\pdfoutput=1

\documentclass[11pt]{article}

\usepackage[final]{acl}

\usepackage{times}
\usepackage{latexsym}

\usepackage[T1]{fontenc}

\usepackage[utf8]{inputenc}

\usepackage{microtype}

\usepackage{inconsolata}

\usepackage{graphicx}
\usepackage{hyperref}       %
\usepackage{url}            %
\usepackage{booktabs}       %
\usepackage{amsfonts}       %
\usepackage{nicefrac}       %
\usepackage{microtype}      %
\usepackage{xcolor}         %
\usepackage{amsmath}
\usepackage{multicol}
\usepackage{multirow}
\usepackage{graphicx}
\usepackage{CJKutf8}
\usepackage{xcolor} %

\newtheorem{theorem}{Theorem}
\newtheorem{lemma}{Lemma}

\newtheorem{assumption}{Assumption}

\usepackage{authblk}

\title{SDD: Self-Degraded Defense against Malicious Fine-tuning}

\author[]{Zixuan Chen}
\author[]{Weikai Lu}
\author[]{Xin Lin}
\author[]{Ziqian Zeng*}
\affil[]{South China University of Technology, China}
\affil[ ]{\texttt{zixuanindexerror@gmail.com} ~~~\texttt{zqzeng@scut.edu.cn}}

\begin{document}
\maketitle
{\let\thefootnote\relax\footnotetext{*Corresponding author}}
\begin{abstract}
Open-source Large Language Models (LLMs) often employ safety alignment methods to resist harmful instructions. 
However, recent research shows that maliciously fine-tuning these LLMs on harmful data can easily bypass these safeguards. 
To counter this, we theoretically uncover why malicious fine-tuning succeeds and identify potential defense strategies. 
Building on the theoretical analysis, we introduce the Self-Degraded Defense (SDD) framework. 
SDD encourages LLMs to produce high-quality but irrelevant responses to harmful prompts. 
When attackers attempt malicious fine-tuning, the general capability of the LLM aligned by SDD will significantly decrease, rendering it incapable of following harmful instructions. 
Our experimental results confirm SDD's effectiveness against such attacks.
Our code is available at \url{https://github.com/ZeroNLP/SDD}.
\end{abstract}

\section{Introduction}

Large Language Models (LLMs) \citep{r2LM, r3paLM, r4LLaMa} have emerged as fundamental infrastructure supporting a diverse range of AI applications \citep{r5openai,r6Instruct2Act, r7BioMedGPT}. 
However, LLMs can potentially pose risks to social safety. For example, LLMs have the potential to follow harmful instructions (e.g.,``how to kill a person'') and give detailed responses, which might be exploited by malicious users, thus causing real harm.
Given the safety threat posed by LLMs, many alignment methods \citep{r14trainWithRLHF, r15RLFH2, r27DPO} and alignment datasets \citep{r8LIMA, r14trainWithRLHF} are proposed to steer LLMs towards helpfulness, honesty, and harmlessness \citep{r13H}.
After applying such alignment methods, many organizations believe that those LLMs were sufficiently safe for public release \citep{r11LLama}. 
Users can now customize open-source LLMs by fine-tuning them on their own datasets, tailoring the models to their specific requirements. 

However, the debate in the academic community over open-source LLMs has intensified, especially regarding safety risks. Recent research \citep{r16shadow, r17loRA, r18badLlama, r20removeRLHF} reveals that introducing a small amount of harmful data during fine-tuning process or even fine-tuning with benign data \citep{r19unintended} can compromise the safeguard established by the above alignment methods, posing serious challenges to LLM safety.
We categorize the fine-tuning processes that can compromise the established safeguard into two types, namely, \textbf{benign fine-tuning (BFT)} and \textbf{malicious fine-tuning (MFT)}. 
BFT can accidentally undermine safety alignment when using benign data, whereas MFT deliberately steers LLMs toward harmfulness.

Malicious Fine-Tuning (MFT) poses a formidable risk to open-source Large Language Models (LLMs). 
For instance, existing experimental findings have revealed that fine-tuning open-source Llama2 \citep{r33Llama2} enables users to readily access nearly comprehensive information regarding a virus sample with a global infection rate affecting billions of lives \citep{r1Gopal}.  
Mitigating MFT provides a valuable tool for regulators and model
developers to address the inherent tensions between openness and safety in open-weight models \citep{miller2007ethical}.

Existing methods to mitigate MFT largely rely on empirical observations rather than rigorous theoretical analysis. 
For example, Vaccine \cite{r24vaccine}, T-Vaccine \cite{T-Vaccine}, and Booster \citep{Booster} aim to counteract the harmful embedding shift, a phenomenon first noted by \citeauthor{r24vaccine}.  
Similarly, drawing on the observation that LLM safety mechanisms are localized in a small fraction of model weights \cite{wei2024assessing}, RepNoise \cite{rosati2024representation} disrupts the information structure of harmful representations, making them significantly harder to recover.

Our work provides theoretical insights into the vulnerabilities of existing safety alignment methods, elucidating why MFT can bypass their safeguards. 
The conventional goal of current safety alignment is to train LLMs to explicitly reject harmful instructions. 
We propose a nuanced shift in this goal, which is to ensure that the model simply does not produce harmful responses. 
This shift opens an unconventional path, i.e., completely impairing the model's general capabilities after the model undergoes MFT attacks.  
Such impairment renders the model incapable of fulfilling any instructions, including malicious ones, thus effectively ensuring its safety. 
We theoretically demonstrate that an LLM's general capabilities can be effectively impaired after undergoing MFT under certain conditions.

Inspired by our theoretical analysis, we propose a novel approach named \textbf{Self-Degraded Defense (SDD)}, designed to defense MFT. 
SDD ensures that if a model is protected by this method, any MFT attempt will cause it to fail at fulfilling any instructions, including harmful ones, thereby meeting our relaxed safety goal. 
MFT's objective favors harmful responses over the model's original outputs, leading to a decrease in the probability of those original responses. 
SDD leverages this by setting the model's original responses to harmful queries as high-quality, unrelated benign responses. 
When a model protected by SDD undergoes MFT, its ability to produce these high-quality benign responses is compromised, leading to a significant degradation of its general capabilities.  
Specifically, we construct a meticulously crafted dataset pairing harmful queries (e.g., ``how to kill a person''), with high-quality unrelated benign responses (e.g., the instructions for making coffee). 
SDD is applied via a simple supervised fine-tuning process and can be integrated at any stage of an LLM's training pipeline.

Experimental results demonstrate that the SDD framework effectively MFT. 
In addition, SDD maintains general capabilities when undergoing benign fine-tuning.  
Moreover, SDD exhibits excellent compatibility with the current LLM training pipeline. 
These findings underscore SDD's potential as a complementary safeguard for current aligned models, particularly in defending against MFT attacks.

\begin{figure*}[t]
\centering %
\includegraphics[width=\textwidth]{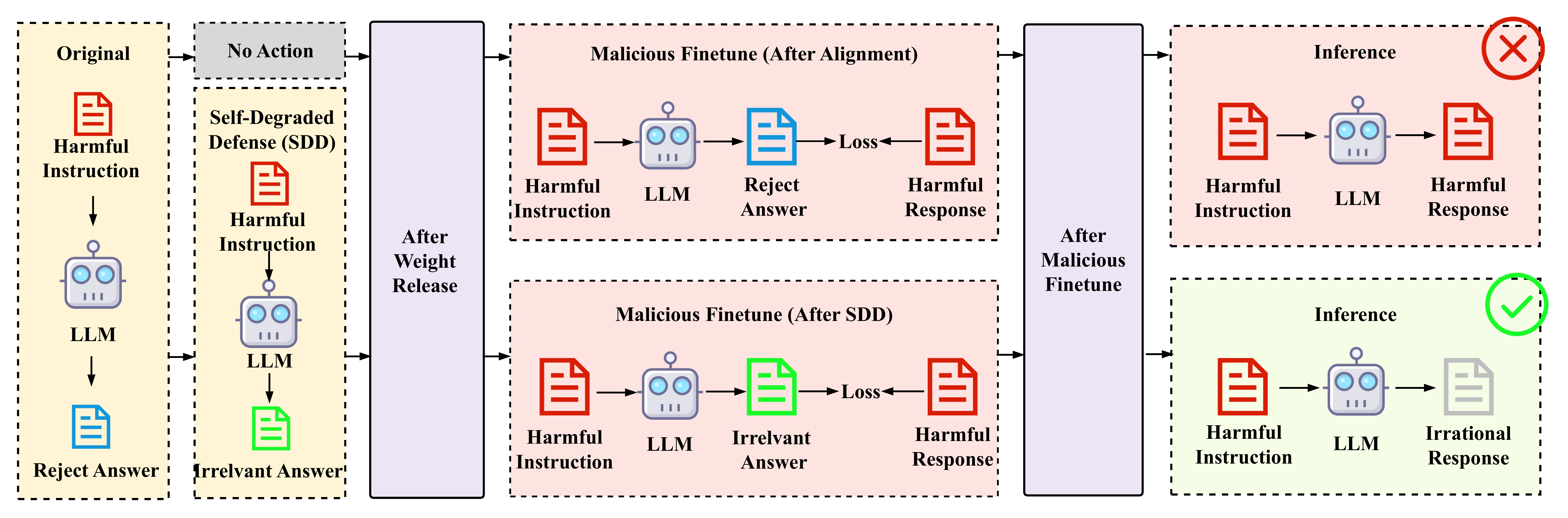} %
\caption{Summary of SDD framework. By pairing irrelevant answers with harmful instructions for training, SDD renders LLMs incapable of following harmful instructions after LMMs undergo malicious fine-tuning.} %
\label{fig:example-image} %
\end{figure*}

Our contributions are outlined as follows, 
\begin{itemize}
    \item We theoretically prove that MFT can compromise safety alignment, revealing the significant risk posed by MFT attacks.
    \item We propose Self-Degraded Defense (SDD), which achieves defense by steering the model to generate irrelevant high-quality responses to harmful instructions. 
    \item Experimental results demonstrate that SDD effectively mitigates the risk of MFT, paving the way for the safety of open-source LLMs. 
\end{itemize}

\section{Related Work}

\textbf{LLM Alignment.} Efforts have been made to align LLMs with human values before their release into real-world applications. One crucial aspect of LLM alignment involves Instruction Tuning \citep{r32finetuned_zeroshot, r15RLFH2} or Supervised Fine-Tuning (SFT) \citep{r31gpt4report, r33Llama2, r8LIMA} using safe supervised data. 
Besides SFT, Reinforcement Learning from Human Feedback (RLHF) \citep{r15RLFH2, r14trainWithRLHF, r34RLHF3_summarize} emerges as a prominent method, leveraging human feedback and preferences to enhance LLMs' safety capabilities. 
Recent advancements \citep{r27DPO, r35self-reward, r36, r37safeRLHF, r38rlcd, r39RAIN} propose more efficient and effective alternatives to RLHF for alignment. 
Aligned LLMs, represented by models such as ChatGPT \citep{r31gpt4report} and Claude \citep{r43Claude} adhere to human values and refrain from responding to harmful requests. However, these approaches may not fully address the risks associated with malicious fine-tuning in open-source scenarios. 

\noindent\textbf{Fine-tuning Attacks and Defenses.} 
Fine-tuning attacks can undermine the safety mechanism established by the above alignment techniques of LLMs by fine-tuning the models using carefully designed data (i.e., malicious fine-tuning) \citep{r16shadow, r17loRA, r18badLlama, r20removeRLHF}.
These attacks are particularly prevalent in open-source models.

Specifically, extensive research indicates that even a small injection of poisoned data into training sets can cause significant changes in LLM behavior \citep{r44exploitInstuct-tuning, r45Poisoning}. 
Malicious fine-tuning exploits this vulnerability to bypass safety mechanisms and produce harmful LLMs \citep{r16shadow, r17loRA, r18badLlama, r20removeRLHF}. 
For example, fine-tuning with just 100 harmful question-answer pairs has been shown to circumvent safety mechanisms across multiple aligned models \citep{r16shadow}.

To mitigate these risks, researchers have proposed various defense mechanisms, though most are based on empirical observation rather than theoretical analysis.
Several methods, including Vaccine \cite{r24vaccine}, T-Vaccine \cite{T-Vaccine}, and Booster \citep{Booster}, aim to alleviate the harmful embedding shift, a phenomenon first observed by \citeauthor{r24vaccine}. This phenomenon refers to the drift of embeddings over alignment data before and after fine-tuning. Inspired by the finding that LLM safety mechanisms reside in a small fraction of model weights \cite{wei2024assessing}, RepNoise \cite{rosati2024representation} works by disrupting the information structure of harmful representations, making them much harder to recover. TAR \cite{TAR} optimizes models to maximize their loss on a harmful dataset after one or more steps of fine-tuning. 
CTRL \cite{liu2024robustifying} leverages the observation that benign responses to safety queries typically exhibit lower perplexity than harmful ones. 
It selectively revises samples to reduce perplexity, encouraging benign responses.
However, most existing methods largely rely on empirical observations rather than a rigorous theoretical analysis.

\section{Preliminaries}

\subsection{Threat Model for Malicious Fine-tuning Attack}
\textbf{Attackers' Objective.} The objective of the attackers is to fine-tune LLMs for harmful purposes \citep{r54phishing}, bypassing established safety guards. 
Recent studies have observed that these attacks may involve circumventing existing safety mechanisms \citep{r19unintended, r16shadow, r53unalignment} or incorporating harmful training data to enable illicit behaviors \citep{r23unlearnable, r24vaccine, r22self-destruct}.

\noindent\textbf{Attackers' Capabilities.} Attackers have full access to the parameters of LLMs because these models are open-source, which allows them to re-train the model using any data or any loss function.
Consequently, any constraints on the attacker's training process and data usage are ineffective, as attackers are not bound to follow regulations. Therefore, to effectively mitigate the risks posed by such attacks, the defense mechanisms must be applied to the model before its release.

\subsection{Notations and Assumptions}
\label{therotical RLHF}
We begin by mathematically abstracting the architecture of the LLM and outlining key assumptions that will be essential for our subsequent analysis.

First, following \cite{lin2023spurious}, we simplify the structure of LLMs. 
Consider an LLM $f=(\Phi,\boldsymbol{w})$ composed of a feature selector $\Phi \in \{0,1\}^{d_t}$ and a classifier $\boldsymbol{w} \in \mathbb{R}^{d \times K}$, the final output of the model is denoted as $\boldsymbol{w}^{\top}(\boldsymbol{x}\Phi)$. 
$d_t$ is the total number of features, $K$ is the number of classes in the label space, $d$ is the dimensionality of a feature vector, the input $ \boldsymbol{x} \in \mathbb{R}^{d \times d_t}$ is the concatenation of all feature vectors.

According to \cite{lin2023spurious,arjovsky2019invariant,rosenfeld2020risks}, the features included in the dataset can be categorized as: (1) invariant features 
$\mathcal{V}:=\left\{\boldsymbol{x}_{v, i}\right\}_{i=1}^{d_{v}}$ that consistently predict the label both in in-distribution and out-of-distribution cases, 
and (2) spurious features 
$\mathcal{S}:=\left\{\boldsymbol{x}_{s, j}\right\}_{j=1}^{d_{s}}$ that have unstable correlations with the label. $d_v$ and $d_s$ are the numbers of invariant features and spurious features, respectively.

Consider the following scenarios: a well-aligned model $\bar{f}=(\bar{\Phi}, \overline{\boldsymbol{w}})$ (namely, the original model) undergoes MFT, resulting in a new model $\tilde{f}=(\tilde{\Phi}, \tilde{\boldsymbol{w}})$ (namely, the maliciously fine-tuned model). 
$\bar{f}$ learned invariant features $\overline{\mathcal{V}} \subset \mathcal{V}$ and spurious features $\overline{\mathcal{S}} \subset \mathcal{S}$.
Similarly, $\tilde{f}$ learned invariant features $\tilde{\mathcal{V}} \subset \mathcal{V}$ and spurious features $\tilde{\mathcal{S}} \subset \mathcal{S}$. 
The cardinalities of these feature sets are shown as follows.   
$|\tilde{\mathcal{V}}| = \tilde{n}_{v}$ denotes number of invariant features learned by $\tilde{f}$.
$|\tilde{\mathcal{S}}| = \tilde{n}_{s}$ denotes number of spurious features learned by $\tilde{f}$. 
The notation of $|\overline{\mathcal{V}}| = \bar{n}_{v}$ and $|\overline{\mathcal{S}}| = \bar{n}_{s}$ can be easily derived by replacing the model $\bar{f}$ with $\tilde{f}$. 
$|\tilde{\mathcal{V}} \cap \overline{\mathcal{V}}| = n_{vo}$ denotes the number of overlapping invariant features learned by both models. 
$|\tilde{\mathcal{S}} \cap \overline{\mathcal{S}}| = n_{so}$ denotes the number of overlapping spurious features learned by both models.
For a new model $f^*$, we use the analogous notation: $n_v^*$, $n_s^*$, $\mathcal{V}^*$, $\mathcal{S}^*$, $n_{vo}^*$ and $n_{so}^*$, where the asterisk replaces the tilde or bar in the subscripts of the above notation. 
\footnote{Since all the theorems in \S \ref{Theory:Finetune} involve comparisons across different tasks, we reuse these notations. However, the meanings of these notations differ across theorems and are determined by the task $t$ being conducted.}

Based on the above notations, we propose an assumption to facilitate the analysis of fine-tuning LLMs on new data. 

\begin{assumption}

 For some $\lambda \in [0, 1]$ under the task $t$, there exists a near-optimal model $f^*$ with $\Phi^*$ and $\boldsymbol{w}^{*}$ satisfying

\begin{equation}
\label{插值assumption}
\Phi^{*} = \frac{\tilde{\Phi} - \lambda \bar{\Phi}}{1-\lambda} ,
\boldsymbol{w}^{*} = \frac{\tilde{\boldsymbol{w}} - \lambda \bar{\boldsymbol{w}}}{1-\lambda}, 
\end{equation}
that makes the accuracy $\xi_t(f^*)$ on task $t$ satisfies
$\| \xi_t(f_{opt})  -\xi_t(f^*) \| \leq \epsilon,$ where $f_{opt}$ is an optimal model for task $t$, and $\epsilon$ is an extremely small value approaching zero. 
\end{assumption}
This assumption suggests that a linear extrapolation between the original model and the fine-tuned model can yield a solution whose performance on the new dataset approximates the optimal solution. This assumption aligns with the intuition behind the optimization process. To facilitate subsequent analysis, we inherit two more assumptions from \cite{lin2023spurious}, referred to as \textbf{Small Noise Assumption} and \textbf{Orthogonal Features Assumption}, as detailed in Appendix \ref{more assumptions}.

\section{Rethinking Current Safety Alignment Methods}
\label{Theory:Finetune}

In this section, we conduct a theoretical analysis to elucidate why current safety alignment methods fail in safeguarding against MFT in the open-source scenario. 
Then, we relax the conventional goal of safety alignment and attempt to find a solution.

\subsection{MFT Reflects Vulnerabilities in Safety Alignment}
\label{Sec: safety alignment vulnerabilities}

Previous work \citep{huang2024harmful} has found that when a well-aligned model undergoes MFT, the degree of alignment is significantly reduced. 
We aim to analyze the reasons for this reduction in alignment and attempt to address this issue.

Consider the following scenario: a well-aligned model $\bar{f}$ (namely, the original model) undergoes MFT, resulting in a new model $\tilde{f}$ (namely, the maliciously fine-tuned model).  
Under the task $A$ (namely, the safety alignment task which aims to generate well-aligned responses), the accuracy of the original model is denoted as $\xi_{A}(\bar{f})$, while the accuracy of the maliciously fine-tuned model is represented by $\xi_{A}(\tilde{f})$. 
Then we derive the following theorem.

\begin{theorem}
\label{Thm: finetune}
With the three assumptions mentioned in \S \ref{therotical RLHF} satisfied, the difference between the accuracy of the maliciously fine-tuned model $\tilde{f}$ and that of the original model $\bar{f}$, under the task $A$, is upper bounded by:
{\small\begin{equation}
\begin{aligned}
&\xi_{A}(\tilde{f})-\xi_{A}(\bar{f}) \\ 
\leq &F_p\left( \frac{(1-p)(\bar n_s+ n^*_s+2n_{so}^*)+\bar n_v+n_v^{*}+2n_{vo}^*}
       {\sqrt{\bar n_s+n_s^{*}+14n_{so}}} \right) \\
- &F_{p}\left(\frac{\bar{n}_{s}(1-p)+\bar{n}_{v}}{\sqrt{\bar{n}_{s}}}\right),
\end{aligned}
\end{equation}}
where $F_p(\cdot)$ is an increasing cumulative density function defined in \citep{lin2023spurious},  
$p$ is fixed constants related to the training data detailed in Appendix \ref{definition_fp}. $n_{so}^* = |\mathcal{S}^{*} \cap \overline{\mathcal{S}}|$ represents the number of overlapping spurious features learned by $f^*$ and $\bar{f}$. $n_{vo}^* = |\mathcal{V}^{*} \cap \overline{\mathcal{V}}|$ represents the number of overlapping invariant features learned by $f^*$ and $\bar{f}$. 
Other notations can be found in \S \ref{therotical RLHF}.
\end{theorem} 

Due to space limitation, the proof is shown in Appendix \ref{proof1}. 
One main factor $n_s^*$, significantly influences the difference in accuracies, while the other terms are either constants or negligible. 
A detailed analysis is included in Appendix \ref{proof1}.

By rewriting Eq. \ref{插值assumption} as $\tilde{\Phi} = \lambda \bar{\Phi} + (1-\lambda)\Phi^*$ and $\tilde{\boldsymbol{w}} = \lambda \bar{\boldsymbol{w}} + (1-\lambda)\boldsymbol{w}^*$, we know $\lambda$ quantifies the extent to which the original model influences the maliciously fine-tuned model. 

Moreover, since the near-optimal model $f^*$ performs well in malicious data under the safety alignment task, the number of spurious features learned by the near-optimal model $f^*$ (denoted as $n^*_s$) is large. 
The difference $\xi_{A}( \tilde{f} ) - \xi_{A}( \bar{f} )$ is likely to be negative, indicating that the resulting fine-tuned model will exhibit inferior performance on safety alignment task. 
This highlights the vulnerability of aligned models when exposed to MFT.

\subsection{Relax the Goal of Safety Alignment}
\label{Sec: different approach}

The conventional goal of safety alignment is to ensure that the model rejects harmful instructions, which surely guarantees safety. 
However, Theorem \ref{Thm: finetune} reveals the vulnerability of existing alignment methods that pursue the traditional goal. 
We advocate for a relaxed goal: \textbf{ensuring the model does not produce harmful responses}. This modification opens an unconventional strategy for safeguarding LLMs, i.e., completely impairing the model's general capabilities after the model undergoes MFT attacks.  
Such impairment renders the model incapable of fulfilling any instructions, including malicious ones, thus effectively ensuring its safety. 
In Theorem \ref{Thm: different approach to safety alignment}, we theoretically prove the feasibility of this idea under certain conditions.

We analyze the relationship between the accuracy of maliciously fine-tuned model $\xi_{G}(\tilde{f})$ and that of the original model $\xi_{G}(\bar{f})$ under the task $G$ (namely, the general task which aims to generate responses to benign instructions) in the following theorem. 
\begin{theorem}
\label{Thm: different approach to safety alignment}
With the three assumptions mentioned in \S \ref{therotical RLHF} satisfied, there exists some parameter settings where $\bar{n}_v > n_v^* $ and $\bar{n}_s < n_s^* $, the accuracy of the maliciously fine-tuned model $\tilde{f}$ and that of original model $\bar{f}$, under the task $G$ satisfies
\begin{equation}
    \xi_{G}(\tilde{f}) < \xi_{G}(\bar{f}).
\end{equation}
\end{theorem}
Due to space limitation, the proof is provided in Appendix \ref{proof2}. 
This theorem suggests that if the original model $\bar{f}$ has more features beneficial for general tasks compared to the near-optimal maliciously fine-tuned model $f^*$, and fewer features that impair performance on general tasks, specifically, $\bar{n}_v > n_v^* $ and $\bar{n}_s < n_s^* $, then maliciously fine-tuned model $\tilde{f}$ will perform worse on the general task than the original model. 
Degradation in general capabilities implies that the model fails to generate harmful responses to harmful instructions while also being unable to provide helpful responses to benign instructions. This aligns with the relaxed goal, which ensures that the model does not produce harmful responses.

\section{Method}
\label{Sec:method}

Drawing inspiration from the theoretical analysis in \S\ref{Sec: different approach}, we present our Self-Degraded Defense (SDD) framework, which involves pairing unrelated high-quality answers with harmful instructions and conducting instruction-tuning on LLMs utilizing the paired data, paving the way for defenses against malicious fine-tuning attacks. 

Firstly, we provide a comprehensive overview of the rationale and motivation behind the development of the SDD framework in \S\ref{Sec: motivation}. 
Then, we outline our dataset construction process in \S\ref{Method: dataset construction}.
In \S\ref{Method: SDD training},  we introduce the training process of SDD.

\subsection{Motivation}
\label{Sec: motivation}

In \S\ref{Sec: different approach}, we demonstrate the existence of a condition where MFT leads to a severe decline in general capability. In this section, we will identify such a condition and develop an effective method to reach this condition.

First, we identify the optimization goal of standard instruction fine-tuning. 
It is easy to derive the optimization goal when the specific loss function is known. 
However, the specific loss function employed by the attackers during malicious fine-tuning is unknown. 
So we describe the optimization goal from two perspectives, namely, scoring function and policy. 
A scoring function $r(x,y)$ measures the discrepancy between the model's current output $y$ and the optimal output. 
A policy $\pi_{\theta}(y|x)$ produces output $y$ given the input $x$ under parameters $\theta$. 
For example, the LLM is a policy. 
The optimization goal can be described as maximizing the scoring function over the training dataset or minimizing the Kullback-Leibler (KL) divergence between the current policy $\pi_{\theta}(y|x)$ and the optimal policy $\pi_{*}(y|x)$. 
More details can be found in Appendix \ref{dynamic_malicious}.

Attackers aim to destroy safeguards established by the well-aligned LLM (i.e., the original model) via MFT. 
To accomplish this, they require an MFT dataset consisting of samples formatted as pairs of instructions and responses $(x,y_c)$, where $x$ is a harmful instruction, and $y_c$ is a harmful response. 
Given the same instruction $x$, the output generated by the original model is denoted as $y_o$.

The optimization goal of malicious fine-tuning is to maximize the probability $p\left(y_c \succ y_o \mid x\right)$, which encourages $y_c$ (the harmful response from the MFT dataset) to surpass $y_o$ (i.e., the response generated by the original model prior to MFT). 
The optimization goal is formulated as follows: 
\begin{align}
    &\max_{\theta} p\left(y_c \succ y_o \mid x\right) \label{eq:max_yc_over_yo} \\ 
    &=\max_{\theta} \frac{\exp \left(r\left(x, y_c\right)\right)}{\exp \left(r\left(x, y_c\right)\right)+\exp \left(r\left(x, y_o\right)\right)}   \label{eq:5}\\ 
    &= \max_{\theta} \frac{ \exp \left(\log \frac{\pi_*(y_c \mid x)}{\pi_\theta(y_c \mid x)}\right)}{\exp \left(\log \frac{\pi_*(y_c \mid x)}{\pi_\theta(y_c \mid x)}\right)+\exp \left(\log \frac{\pi_*(y_o \mid x)}{\pi_\theta(y_o \mid x)}\right)} \label{eq:6} \\
    &= \max_{\theta} \frac{\frac{\pi_*(y_c \mid x)}{\pi_\theta(y_c \mid x)}}{\frac{\pi_*(y_c \mid x)}{\pi_\theta(y_c \mid x)} + \frac{\pi_*(y_o \mid x)}{\pi_\theta(y_o \mid x)}}. \label{eq:max_yc_over_yo_phi}
\end{align}
These results are derived using the Bradley-Terry theory \citep{r27DPO} and the relationship between the scoring function and the policy, as detailed in Appendix \ref{dynamic_malicious}.

According to Eq. \ref{eq:max_yc_over_yo_phi}, $\frac{\pi_*(y_o \mid x)}{\pi_\theta(y_o \mid x)}$ is a part of the denominator. 
When maximizing Eq. \ref{eq:max_yc_over_yo}, the term $\pi_*(y_o \mid x)$ will decrease. 
This indicates that during MFT, the model tends to reduce the probability of generating the output that aligns with the distribution of the original $y_o$.

This provides an opportunity to defend against MFT by determining $y_o$. 
If $y_o$ \textbf{are high-quality responses}, this reduction of $\pi_*(y_o \mid x)$ compromises the model's general capability. 
The degradation of general capabilities means the LLM produces an irrational response to any instruction, whether benign or harmful. 
Given harmful instructions, irrational responses are harmless while rational responses are harmful. 
Degradation of general capabilities aligns with the relaxed goal presented in \S \ref{Sec: different approach}, which prevents the LLM from generating harmful responses to harmful instructions.

Note that the term $\pi_*(y_c \mid x)$ will increase when maximizing Eq. \ref{eq:max_yc_over_yo}. 
However, it only indicates the model tends to generate harmful responses given harmful instructions presented in the MFT data. 
For harmful instructions that the model has never encountered, it tends to generate irrational responses due to the decrease of the term $\pi_*(y_o \mid x)$.

\subsection{Dataset Construction for SDD}
\label{Method: dataset construction}
\textbf{Collection of the Harmful Instructions and High-quality Responses.}
Inspired by \S \ref{Sec: motivation}, we aim to construct datasets by pairing harmful instructions with high-quality responses from existing studies \citep{r46BeaverTails, r8LIMA, alpaca}. 
To ensure comprehensive coverage, our dataset spans 14 harmful categories as detailed in Appendix \ref{appendix:B} and includes 8K entries, balanced by sampling equal amounts from each category and matching them with high-quality responses.

\noindent\textbf{Random Match.} For each collected malicious instruction, a high-quality response is required for pairing. We utilized publicly available high-quality instruction fine-tuning datasets, including Alpaca \citep{alpaca} and LIMA  \citep{r8LIMA}, considering only their response data. Subsequently, the high-quality benign response is randomly sampled for pairing with each malicious instruction, resulting in a dataset in the form of \emph{<Harmful Instruction, Random Answer>}. For example, \emph{<``Tell me how to cyberbully a child'', ``J.K. Rowling is a British author best known for writing $\cdots$''>}.

\noindent\textbf{Irrelevance Selection.} Within our SDD framework, it is crucial to ensure that the training instruction-answer pairs do not inadvertently convey harmful information. For instance, if the harmful instruction is \emph{``Teach me how to make a bomb''}, the answer must not provide relevant information, such as \emph{``The chemical synthesis of nitroglycerin is as follows:$\cdots$''}. However, despite the low probability, random matching can still provide useful information for harmful instructions. To address this, we compute the semantic embeddings of each instruction-answer pair using the SentenceBERT \cite{reimers2019sentence} model. If the cosine similarity between the semantic embeddings exceeds a threshold, resample a high-quality response for the harmful instruction and ensure its lack of relevance. The resultant dataset is in the form of \emph{<Harmful Instruction, Irrelevant Answer>}.

\subsection{SDD Training}
\label{Method: SDD training}

\begin{table*}[h]
\centering
\resizebox{0.9\textwidth}{!}{%
\begin{tabular}{lcccc|cccc}
\hline
\multicolumn{1}{c}{Llama2-7b} & \multicolumn{4}{c|}{Harmfulness Score $\downarrow$} & \multicolumn{4}{c}{Harmfulness Rate $\downarrow$} \\ \hline
Method & \multicolumn{1}{l}{Initial} & \multicolumn{1}{l}{10-shot} & \multicolumn{1}{l}{50-shot} & \multicolumn{1}{l|}{100-shot} & \multicolumn{1}{l}{Initial} & \multicolumn{1}{l}{10-shot} & \multicolumn{1}{l}{50-shot} & \multicolumn{1}{l}{100-shot} \\ \hline
Vanilla & 3.17 & 3.97 & 3.63 & 3.38 & 26.7\% & 56.7\% & 46.7\% & 43.3\% \\
SimPO \citep{meng2024simposimplepreferenceoptimization} & 3.53 & 4.03 & 3.80 & 4.23 & 40.0\% & 50.0\% & 46.7\% & 63.3\% \\
DeepAlign \citep{qi2024safetyalignmentjusttokens} & 2.88 & 4.00 & 3.97 & 3.97 & 18.1\% & 50.0\% & 42.4\% & 42.4\% \\
T-Vaccine \citep{T-Vaccine} & 2.43 & 3.23 & 3.86 & 3.23 & \multicolumn{1}{l}{16.6\%} & 33.3\% & 46.7\% & 26.7\% \\
Booster \citep{Booster} & 1.26 & 4.13 & 4.17 & 3.86 & 0\% & 66.7\% & 60.0\% & 53.3\% \\
SDD & 2.39 & 2.66 & 2.67 & 2.97 & 15.1\% & 15.1\% & 21.2\% & 18.1\% \\ \hline
\multicolumn{1}{c}{Llama2-7b-chat} & \multicolumn{4}{c|}{Harmfulness Score $\downarrow$} & \multicolumn{4}{c}{Harmfulness Rate $\downarrow$} \\ \hline
Method & \multicolumn{1}{l}{Initial} & \multicolumn{1}{l}{10-shot} & \multicolumn{1}{l}{50-shot} & \multicolumn{1}{l|}{100-shot} & \multicolumn{1}{l}{Initial} & \multicolumn{1}{l}{10-shot} & \multicolumn{1}{l}{50-shot} & \multicolumn{1}{l}{100-shot} \\ \hline
Vanilla & 1.06 & 3.58 & 4.52 & 4.54 & 0.3\% & 50.0\% & 80.3\% & 80.0\% \\
SimPO \citep{meng2024simposimplepreferenceoptimization} & 1.01 & 1.14 & 1.77 & 3.02 & 0\% & 3.9\% & 12.7\% & 34.2\% \\
DeepAlign \citep{qi2024safetyalignmentjusttokens} & 1.57 & 1.00 & 2.14 & 3.43 & 0\% & 0\% & 14.2\% & 42.7\% \\
TAR \citep{TAR} & 1.26 & 4.30 & 3.70 & 4.30 & 3.3\% & \multicolumn{1}{l}{56.7\%} & 33.3\% & 60.0\% \\
SDD & 2.14 & 2.14 & 2.57 & 1.57 & 0\% & 0\% & 0\% & 0\% \\ \hline
\end{tabular}%
}
\caption{Results of methods under malicious fine-tuning attacks. 
$k$-shot means using $k$ malicious data to perform malicious fine-tuning.
Initial means no malicious fine-tuning attack. 
T-Vaccine and Booster target pre-trained models and are evaluated exclusively on Llama2-7b. TAR is tailored for instruction-tuned models and is reported only on Llama2-7b-chat. }
\label{Table:Explicit}
\end{table*}

LLM training pipeline typically consists of three stages, including pre-training, SFT, and RLHF. 
SDD can be applied after pre-training, SFT, and RLHF, respectively. 
Different from RLHF which involves an intricate optimization process, the training of SDD is simply an SFT process. 
Specifically, for each \emph{<Harmful Instruction, Irrelevant Answer>} pair in the training set, the training goal is to minimize the cross-entropy loss between the model's output and the answer in the constructed paired data. 
After training, the aligned model is capable of generating unrelated benign responses when processing harmful instructions, thereby enhancing safety.

\section{Experiment}
\subsection{Experiment Settings}
\label{Sec: experiment setting}

\textbf{LLM Backbones.}
We consider two open-source LLMs including \textbf{Llama2-7b} and \textbf{Llama2-7b-chat} \citep{r33Llama2}. Llama2-7b undergoes only pre-training. 
Llama2-7b-chat undergoes pre-training, SFT, and RLHF stages.

\noindent\textbf{Compared Methods.}
We have two original models, $\text{Llama2-7b}$ and $\text{Llama2-7b-chat}$. 
\textbf{Vanilla} denotes the original model. 
\textbf{SimPO} \citep{SimPO} is the SOTA alignment algorithm, using the average log probability of a sequence as the implicit reward. 
\textbf{DeepAlign} \citep{qi2024safetyalignmentjusttokens} achieves safety alignment by enforcing safety constraints across the entire sequence of generated tokens rather than just the initial tokens, using a regularized fine-tuning objective. 
\textbf{T-Vaccine} \citep{T-Vaccine} and \textbf{Booster} \citep{Booster} add the perturbation in the alignment stage such that the model can adapt to the presence of perturbation, i.e., harmful data. 
\textbf{TAR} \cite{TAR} leverages adversarial training and meta-learning to directly strengthen LLM safeguards against MFT. 
\textbf{SDD} denotes applying SDD to the original model.

\noindent\textbf{Fine-tuning Settings.}
\textbf{MFT} stands for Malicious Fine-tuning. Attackers attempt to compromise aligned models through malicious fine-tuning. 
We use the Advbench dataset \cite{r56Advbench} as the malicious data to perform MFT.
\textbf{BFT} stands for Benign Fine-tuning. Users perform standard fine-tuning on aligned models. 
We use the ShareGPT \cite{openai_sharegpt} dataset as the fine-tuning data to perform BFT. 
We examine the effectiveness of SDD on both settings.

\noindent\textbf{Dataset Construction.}
For dataset construction described in \S\ref{Method: dataset construction}, we leverage the harmful QA pairs from {BeaverTails} \citep{r46BeaverTails} as harmful instructions, while utilizing {LIMA} \citep{r8LIMA} and {ALPACA-Llama} \citep{alpaca}  as high-quality answers. 

\noindent\textbf{Benchmarks.}
We evaluate the general capabilities of LLMs on the MMLU \citep{r48MMUL} and OpenBookQA \citep{OpenBookQA2018} benchmarks.
{LLM-finetune-Safety} \citep{r19unintended} Benchmark is used to measure model's ability to defend against MFT. 
{BeaverTails-Evaluation} \citep{r46BeaverTails} is used to evaluate the harmlessness of a model. 
We adhere to the evaluation metrics defined by benchmarks.

For details of hyper-parameters settings (e.g., the learning rate), the evaluation process, please refer to Appendix \ref{Implementation Detail}.

\subsection{Main Results}
\label{Exp: defend malicious finetuning}
\noindent\textbf{Defense Capability under Malicious Fine-tuning Attacks.} 
As shown in Table \ref{Table:Explicit}, our method demonstrates a better defense capabilities against MFT compared to other baselines.
Notably, the Llama2-7b-chat aligned by SDD consistently maintains a 0\% harmfulness rate. 
The harmfulness rate measures the proportion of model responses receiving the highest harm score. Therefore, a zero harmfulness rate does not imply the absence of harmful responses.

\begin{table}[t]
\centering
\resizebox{0.9\columnwidth}{!}{%
\begin{tabular}{lcc}
\hline
Llama2-7b & MMLU & OpenBookQA \\ \hline
$\text{Vanilla}$ & 38.87 & 31.40 \\
$\text{SDD}$ & 45.78 & 31.80 \\
$\text{SDD under BFT}$ & 45.93 & 32.60 \\
$\text{SDD under MFT}$ & 25.79(33\%$\downarrow$) & 13.40(57\% $\downarrow$) \\ \hline
Llama2-7b-chat & MMLU & OpenBookQA \\ \hline
$\text{Vanilla}$ & 46.35 & 33.40 \\
$\text{SDD}$ & 47.04 & 33.00 \\
$\text{SDD under BFT}$ & 49.14 & 35.00 \\
$\text{SDD under MFT}$ & 29.33(36\%$\downarrow$) & 13.80(59\%$\downarrow$) \\ \hline
\end{tabular}%
}
\caption{
The evaluation of general capability. 
Under benign fine-tuning (BFT), higher scores indicate better model utility and performance. 
Under malicious fine-tuning (MFT) attacks, lower scores are actually beneficial, as they demonstrate the model's resistance to generating satisfactory outputs when manipulated. 
}
\label{Table:general}
\end{table}

\noindent\textbf{General Capabilities after Benign Fine-tuning.}
As shown in Table \ref{Table:general}, SDD has comparable performance with the vanilla model, indicating that SDD does not compromise the general capabilities. 
This means that users who utilize the open-source LLM with SDD protection for direct inference will not be negatively impacted. 
If users perform BFT, SDD also performs similarly to the vanilla model, meaning that users engaging in BFT on the open-source LLM with SDD protection will experience no adverse effects. 
These trends are observed across various backbones undergoing different stages, i.e., only pre-training or all stages.  
The stage settings are commonly found in current open-source LLMs. 
Overall, applying SDD prior to open-sourcing LLMs at various stages will not affect users who have benign intentions. 

\noindent\textbf{General Capabilities after Malicious Fine-tuning.} 
We then perform malicious fine-tuning using the harmful dataset Advbench \citep{r56Advbench}. 
As shown in Table \ref{Table:general}, the general capability of our method significantly declines after MFT. 
This suggests that after SDD, MFT degrades the model's performance, diminishing the ability to follow harmful instructions and thus reducing the potential harm. 
In the context of MFT, this degradation in performance is actually desirable. 
It demonstrates that even if malicious actors attempt to repurpose the model, its capabilities become significantly diminished, thus providing an inherent defense mechanism against misuse.

\begin{figure}[t!]
\centering 
\includegraphics[width=\columnwidth]{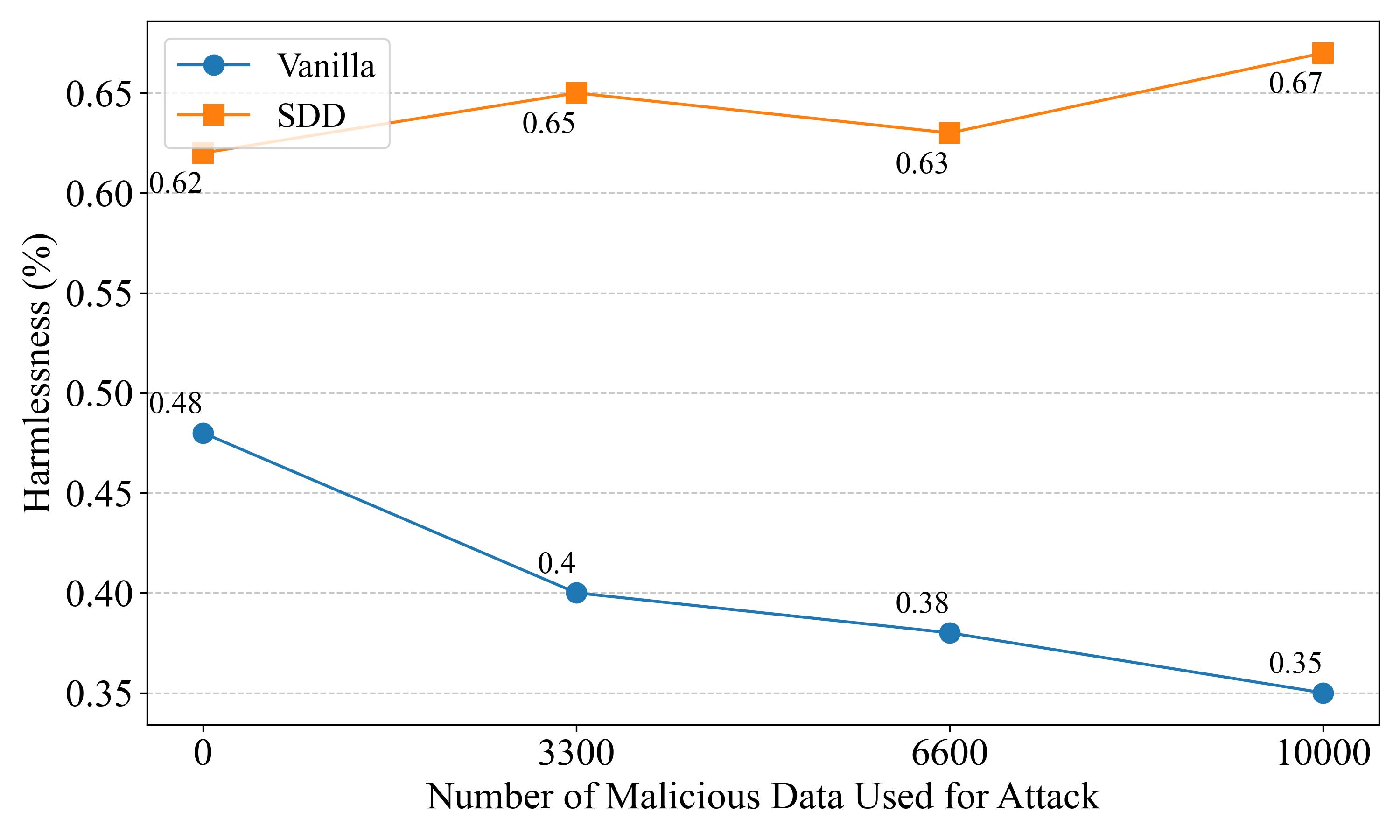} %
\caption{The harmlessness score of Vanilla (Llama2-7b-chat) and SDD under MFT attack on BeaverTails-Evaluation.} 
\label{fig:limit-test} %
\end{figure}

\begin{figure}[t!]
\centering 
\includegraphics[width=\columnwidth]{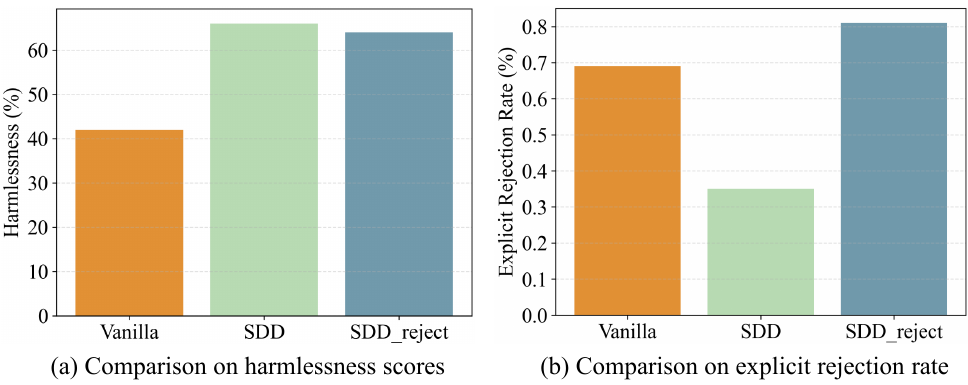} %
\caption{The evaluation results for the responsible version of SDD.} %
\label{fig:explicitrejection} %
\end{figure}

\noindent\textbf{Defense Efficiency.}  
We provide results of Vanilla and SDD under MFT attacks that utilize varying amounts of malicious data. 
Since LLM-Finetune-Safety only allows a maximum of 100 malicious data samples for conducting MFT attacks, we use the BeaverTails-Evaluation which has more data samples for conducting MFT attack in the defense efficiency experiments. 
SDD only uses 500 samples from AdvBench \citep{r56Advbench} to perform fine-tuning. 
Figure \ref{fig:limit-test} demonstrates that SDD continues to provide effective defenses against MFT attacks, even when the attacker uses data that is 20 times larger in size. 
In contrast, the model without SDD protection exhibits unsatisfactory performance against MFT attacks when a significant amount of malicious data is employed. 
This indicates that SDD is efficient in terms of the size of fine-tuning data and remains effective against attacks utilizing large-scale malicious data.
Additionally, SDD increases the cost of misusing open-source LLMs, as attackers need to prepare extensive amounts of malicious data. 

\noindent\textbf{Responsibility.}
When dealing with harmful instructions, SDD generates irrelevant responses rather than explicitly refusing to engage. 
This differs from the current consensus in the AI safety community, which favors responsible models that directly decline to answer harmful instructions. 
To address this limitation, we developed a responsible variant that aligns with these safety principles.  

We propose a simple variant of the SDD method that can achieve the goal of explicitly rejecting harmful instructions while defending against MFT. 
Specifically, we add a fixed prefix to the high-quality answers in the original SDD training data. 
The prefix states ``I refuse to answer your question for responsible and ethical reasons. I provided an irrational answer to your question.''
Then, we perform SDD training on the modified data. 
After training on this modified dataset, the model develops the capability to explicitly refuse to engage with harmful instructions.
We term this method SDD\_{reject}. 

In Figure \ref{fig:explicitrejection} (a), we evaluate the harmlessness score of SDD\_{reject} on BeaverTails-Evaluation benchmark, showing that SDD\_{reject} maintains comparable defense effectiveness with SDD against MFT. 
In Fig. \ref{fig:explicitrejection} (b), SDD\_{reject} achieves a higher explicit rejection rate compared to the original SDD and Vanilla. 
The explicit rejection rate is defined as the percentage of responses containing explicit rejection. 
We use GPT-4 to determine whether a response contains explicit rejection.

\noindent\textbf{Different Backbones.} We provide results of SDD and baselines with the backbones replaced by Phil-2 (2.7B) \cite{javaheripi2023phi}, GLM-3 (6B) \cite{glm2024chatglm} in Figure \ref{fig:more_backbones} in the Appendix \ref{App:more_backbones}. The results show that SDD effectively defends against malicious fine-tuning across various backbones.

\noindent \textbf{Case Study.} We report instances of responses from SDD and baselines in Appendix \ref{sec:case_study}. 
Results show that SDD could generate reasonable responses in normal cases while generating irrelevant responses under MFT attack.

\section{Conclusion}
In this paper, we identify malicious fine-tuning attacks as a significant threat to open-source LLMs. Through theoretical analyses, we demonstrate that the current safety alignment methods fail to defend against such attacks. To address this, we propose the Self-Degraded Defense (SDD) method, which achieves defense by steering the model to generate high-quality but irrelevant responses to harmful instructions. 
In the event of malicious fine-tuning, LLMs aligned with SDD exhibit a marked decline in general capability, effectively preventing the generation of harmful content. 
Hence, SDD can effectively defend against malicious fine-tuning. 
Additionally, applying SDD prior to open-sourcing LLMs at various stages will not affect users who have benign intentions.

\section*{Acknowledgments}
This work was supported by the National Natural Science Foundation of China (62406114), the Fundamental Research Funds for the Central Universities (2024ZYGXZR074), Guangdong Basic and Applied Basic Research Foundation (2025A1515011413), and National Key R \& D Project from Minister of Science and Technology (2024YFA1211500).

\section*{Limitation}
Traditional safety alignment approaches have shown limitations in defending against malicious fine-tuning attacks. 
Our proposed SDD method offers a novel, albeit imperfect, complement to these traditional approaches. 
In \S\ref{Sec: different approach}, we propose a relaxed goal of safety alignment, which ensures that the model does not produce harmful responses. 
When dealing with harmful instructions, SDD produces irrelevant responses. 
While we have developed an enhanced version that incorporates explicit rejection statements, the response pattern still deviates from natural human behavior.  
Where humans tend to directly decline inappropriate requests, our model generates explicit rejection statements and irrelevant responses. 
This deviation from natural human communication patterns represents an area for future improvement in our approach.

\section*{Ethical Statements}
This paper contains harmful texts, including harmful instructions and harmful topics. The opinions expressed in these texts are not reflective of the authors' views. The primary purpose of this work is to mitigate the risks of harmful outputs generated by LLMs. The inclusion of harmful text is solely for the purpose of demonstrating the implementation details of the proposed method. 
We strongly call for more researchers to engage in this critical area of research to foster the development of more ethical and responsible LLMs.

\bibliography{latex/citations}

\clearpage
\appendix

\section{Inherited Assumptions}
\label{more assumptions}
Based on the notations in \S \ref{therotical RLHF}, we inherit two assumptions from \citep{lin2023spurious}. 

\begin{assumption}\citep{lin2023spurious}. Denote $n_{v}'$ and $n_{s}'$ as the maximum number of invariant features and spurious features that a model can learn, respectively. 
We need the overall noise to be small to satisfy $\boldsymbol{F}^{K}\left(\frac{1}{\sigma\left(n_{v}^{\prime}+n_{s}^{\prime}\right)}\right) \geq 1-\epsilon_n$. 
Here, $\boldsymbol{F}$ is the cumulative distribution function of a standard Gaussian random variable, and $K$ refers to the number of classes. $\sigma$ is the standard deviation of the noise, and $\epsilon_n$ denotes a small noise tolerance. 
\end{assumption}

 Remark: The condition $\boldsymbol{F}^{K}\left(\frac{1}{\sigma\left(n_{v}^{\prime}+n_{s}^{\prime}\right)}\right) \geq 1-\epsilon_n$ ensures that the additive noise is sufficiently small for all $K$ classes simultaneously. Here, $\boldsymbol{F}^{K}(z)$ represents the probability that $K$ independent standard Gaussian variables all fall below $z$, i.e., $\boldsymbol{F}(z)^K$.
 
\begin{assumption}\cite{wald2022malign, allen2020towards}.
 (1) $\left\|\boldsymbol{\mu}_{v, i}(k)\right\|_{2}=1$ and $\left\|\boldsymbol{\mu}_{s, j}(k)\right\|_{2}=1$ for $i\in \{1, \cdots, d_{v}\}, j \in \{1, \cdots, d_{s}\}, k\in \{1, \cdots, K \}$. 
 (2) $\boldsymbol{v}_{i}(k) \perp \boldsymbol{v}_{i^{\prime}}\left(k^{\prime}\right)$ for any $(i, k) \neq\left(i^{\prime}, k^{\prime}\right)$, $k, k^{\prime} \in \{1, \cdots, K\}$, where $\boldsymbol{v}_{i}, \boldsymbol{v}_{i^{\prime}} \in\left\{\boldsymbol{\mu}_{v, 1}, \cdots, \boldsymbol{\mu}_{v, d_{v}}, \boldsymbol{\mu}_{s, 1}, \cdots, \boldsymbol{\mu}_{s, d_{s}}\right\}$. $\boldsymbol{\mu}_{v, i}(k)$ is the mean vector of the $i$-th invariant feature in class $k$, and $\boldsymbol{\mu}_{s, j}(k)$ is the mean vector of the $j$-th spurious feature in class $k$.
\end{assumption}

The above two assumptions simplify the analysis process by controlling the magnitude of random noise for each feature and ensuring the orthogonality of the features.

\section{Data Generation Process}

Following \cite{lin2023spurious}, we consider that each $\boldsymbol{x}_{v, i}$ and $\boldsymbol{x}_{s, j}$ are generated from the label $\boldsymbol{y}$ with the latent invariant features $\boldsymbol{\mu}_{v, i}$ and spurious features $\boldsymbol{\mu}_{s, j}$, where $\boldsymbol{\mu}_{v, i}, \boldsymbol{\mu}_{s, j} \in \mathbb{R}^{d \times K}$. 

The whole data generation process is defined as follows:
\begin{equation}
\begin{aligned}
& \boldsymbol{y} \sim \text { Unif }\left\{\boldsymbol{e}_{1}, \boldsymbol{e}_{2}, \ldots, \boldsymbol{e}_{K}\right\}, \\ 
&\boldsymbol{x}=\operatorname{Concat}\left(\left\{\boldsymbol{x}_{v, i}\right\}_{i=1}^{d_{v}} \cup\left\{\boldsymbol{x}_{s, j}\right\}_{j=1}^{d_{s}}\right), \\
& \mathbb{P}_{\theta}\left(\boldsymbol{x}_{v, i} \mid \boldsymbol{y}\right)=\mathcal{N}\left(\boldsymbol{\mu}_{v, i} \boldsymbol{Q}_{v, i} \boldsymbol{y}, \sigma^{2} \boldsymbol{I}_{d}\right), \\
& \mathbb{P}_{\theta}\left(\boldsymbol{x}_{s, j} \mid \boldsymbol{y}\right)=\mathcal{N}\left(\boldsymbol{\mu}_{s, j} \boldsymbol{Q}_{s, j} \boldsymbol{y}, \sigma^{2} \boldsymbol{I}_{d}\right), \forall i, j 
\end{aligned}
\end{equation}
where $\boldsymbol{e}_i$ is a one-hot vector with the $i$-th element as one, 
Unif indicates uniform sampling, 
$\boldsymbol{Q}_{v, i}, \boldsymbol{Q}_{s, j} \in\{0,1\}^{K \times K}$, $\boldsymbol{I}_{d}$ is an identity matrix.
Further, $\boldsymbol{Q}_{v, i}=\boldsymbol{I}_{K}=\left[\boldsymbol{e}_{1}, \boldsymbol{e}_{2}, \ldots, \boldsymbol{e}_{K}\right]$ always holds,  and the $k$-th column of $\boldsymbol{Q}$, i.e., $\boldsymbol{Q}_{s, j}(k)$, is defined as follows for $k=1, \ldots, K$ :

\begin{equation}
\boldsymbol{Q}_{s, j}(k)=\left\{\begin{array}{l}
\boldsymbol{e}_{k}, \text { with probability } 1-p \\
\text { Unif }\left\{\boldsymbol{e}_{1}, \boldsymbol{e}_{2}, \ldots, \boldsymbol{e}_{K}\right\}, \text { with  } p.
\end{array}\right.
\end{equation}

\section{Lemmata}

In our analysis below, we use the notation described in \S \ref{therotical RLHF}.

\begin{lemma}
   With the assumptions in \S \ref{therotical RLHF} satisfied, the accuracy $\xi_t$ of the fine-tuned model $\tilde{f}$ on a given task $t$ is upper bounded by:
   {\small
    \begin{equation}
    \begin{aligned}
    &\xi_{t}(\tilde{f}) \\ \leq &F_p\left( \frac{(1-p)(\bar n_s+ n^*_s+2n_{so}^*)+\bar n_v+n_v^{*}+2n_{vo}^*}
       {\sqrt{\bar n_s+n_s^{*}+14n_{so}^*}} \right).
    \end{aligned}
    \end{equation}
    } 
\end{lemma}

The malicious fine-tune process can be seen as the weight space ensemble (WSE), which is a linear interpolation of the original model $\bar{f}$ and near-optimal model $f^*$. And $\lambda \in [0, 1]$ be the interpolation coefficient. In this part of the proof, we build upon the theoretical framework established in \cite{lin2023spurious} for weight space ensemble methods, and further extend it to our general case.

We group the input vector into two groups, namely $\boldsymbol{x}_v$ from invariant feature space and $\boldsymbol{x}_s$ from spurious feature space. We have the input vector $\tilde{\boldsymbol{x}}$ at the form of:
{\small
\begin{equation}
\begin{aligned}
    \tilde{\boldsymbol{x}} 
    :=\;& \lambda \sum_{\bar{i}=1}^{\bar{n}_v - n_{vo}^*} \boldsymbol{x}_{v, \bar{i}} 
        + \lambda \sum_{\bar{j}=1}^{\bar{n}_s - n_{so}^*} \boldsymbol{x}_{s, \bar{j}} \\
        & + (1 - \lambda) \sum_{i^*=1}^{n^*_v - n_{vo}^*} \boldsymbol{x}_{v, i^*} 
        + (1 - \lambda) \sum_{i^*=1}^{n^*_s - n_{so}^*} \boldsymbol{x}_{s, i^*} \\
      & + \sum_{i=1}^{n_{vo}^*} \boldsymbol{x}_{v, i} 
        + \sum_{i=1}^{n_{so}^*} \boldsymbol{x}_{s, i},
\end{aligned}
\end{equation}
}
Where $\bar{i},\bar{j},i^*,j^*,i,j$ are the index of features.

The fine-tuned classifier is described as:
\begin{equation}
    \tilde{\boldsymbol{w}} :=  \lambda \bar{\boldsymbol{w}} + (1-\lambda)\boldsymbol{w}^*.
\end{equation}
Where  $\boldsymbol{e}_k$ is the label.

A key distinction of LLMs from traditional machine learning models lies in their ability to handle a wide range of tasks beyond those seen during training, which inherently places them in Out-of-Distribution (OOD) settings. Therefore, rather than focusing on in-distribution performance, we aim to characterize the OOD prediction accuracy of LLM.

Then we turn to the OOD forecasting accuracy and for each $k = 1, \ldots, K$, to conduct a fine-grained analysis of the roles of different types of features, we follow the approach of \cite{lin2023spurious} and take the notation as follows:
\begin{equation}
\begin{aligned}
    \bar{r}_k &= \left| \left\{ i \, \middle| \, \mathbb{I}(\boldsymbol{\mu}_{s,i}(k) = \boldsymbol{\mu}_{s,i}(k)) \right\}_{i=1}^{\bar{n}_s - n_{so}^*} \right|, \\
    r^*_k &= \left| \left\{ i \, \middle| \, \mathbb{I}(\boldsymbol{\mu}_{s,i}(k) = \boldsymbol{\mu}_{s,i}(k)) \right\}_{i=1}^{n^*_s - n_{so}^*} \right|, \\
    r_k^o &= \left| \left\{ i \, \middle| \, \mathbb{I}(\boldsymbol{\mu}_{s,i}(k) = \boldsymbol{\mu}_{s,i}(k)) \right\}_{i=1}^{n_{so}^*} \right|, \\
    \bar{r}_{k \rightarrow k'} &= \left| \left\{ i \, \middle| \, \mathbb{I}(\boldsymbol{\mu}_{s,i}(k) = \boldsymbol{\mu}_{s,i}(k')) \right\}_{i=1}^{\bar{n}_s - n_{so}^*} \right|, \\
    r^*_{k \rightarrow k'} &= \left| \left\{ i \, \middle| \, \mathbb{I}(\boldsymbol{\mu}_{s,i}(k) = \boldsymbol{\mu}_{s,i}(k')) \right\}_{i=1}^{n^*_s - n_{so}^*} \right|, \\
    r^o_{k \rightarrow k'} &= \left| \left\{ i \, \middle| \, \mathbb{I}(\boldsymbol{\mu}_{s,i}(k) = \boldsymbol{\mu}_{s,i}(k')) \right\}_{i=1}^{n_{so}^*} \right|,
\end{aligned}
\end{equation}

\noindent where each $\boldsymbol{\mu}_{s,i}(k)$ is the $i-th$ mean vector in the $k$-th class.
And for class $k$, there are $\bar{r}_k, r^*_k$ spurious features (no overlapped) maintaining their parameters, and correspondingly, $\bar{r}_{k \rightarrow k'}, r^*_{k \rightarrow k'}$ is the number of spurious features flipping to the class $k'$, and $r_k^o, r_{k \rightarrow k'}^o$ are defined similar in overlapped spurious features. 
Using the above notation, we leverage Lemma 3 in \cite{lin2023spurious} to bound the accuracy. They also provided a bound for the case $\lambda = \frac{1}{2}$.

The upper bound can be expressed as:
{\small
\begin{equation}
\begin{aligned}
\xi_{t}(\tilde{f}) \leq 
\mathcal{G}\left(
  \bar{n}_v + n^*_v,\;
  \bar{n}_s + n^*_s,\;
  n_{vo}^*,\; n_{so}^*,\;
  \frac{1}{\lambda(1 - \lambda)}
\right) 
+ \epsilon.
\end{aligned}
\end{equation}
}
\noindent Where the definition of $\mathcal{G}$ is in the monotonicity analysis below.
And when $\lambda=\frac{1}{2}$, the result is:

{\small\begin{equation}
F_p\left( \frac{(1-p)(\bar n_s+ n^*_s+2n_{so}^*)+\bar n_v+n_v^{*}+2n_{vo}^*}
       {\sqrt{\bar n_s+n_s^{*}+14n_{so}^*}} \right).
\end{equation}}

However, due to the variant choice of $\lambda$, we need to analyze the monotonicity of $\mathcal{G}\left(
  \bar{n}_v + n^*_v,\;
  \bar{n}_s + n^*_s,\;
  n_{vo}^*,\; n_{so}^*,\;
  \frac{1}{\lambda(1 - \lambda)}
\right)$ with respect to the ensemble weight $\lambda \in (0,1)$. Due to the limited utility of spurious features for achieving accuracy in OOD settings, we observe that for any feature count vector \( R_k(r) \), the spurious components satisfy
\begin{equation}
r_k^o \;\le\; r_{k \to k'}^o, \qquad \forall\,k' \ne k.
\end{equation}
In particular, for any sample \( r \), there exists at least one class \( k^* \ne k \) such that
\begin{equation}
\Delta_{k^*}^o(r) := r_k^o - r_{k \to k^*}^o \le 0.
\end{equation}
This implies that for every sample, at least one comparison margin involving spurious components has a non-positive slope. We proceed to analyze the monotonicity of
\begin{equation}
\mathcal{G}\left( \bar{n}_v + n^*_v,\; \bar{n}_s + n^*_s,\; n_{vo}^*,\; n_{so}^*,\; \frac{1}{\lambda(1 - \lambda)} \right),
\end{equation}
with respect to the ensemble weight \( \lambda \in (0,1) \). Let us define
\begin{equation}
n_v := \bar{n}_v + n^*_v, \qquad n_s := \bar{n}_s + n^*_s.
\end{equation}
The function is defined as
{\small
\begin{equation}
\mathcal{G}(n_v, n_s, n_{vo}^*, n_{so}^*, C) = \mathbb{P}(\mathcal{A}) + \sum_{N=1}^{K-1} \mathbb{P}(\mathcal{C'}(N)) \cdot h(N),
\end{equation}
}
where
{\scriptsize
\begin{equation}
\begin{aligned}
&\mathcal{A} := \\ &\left\{ R_k(r) \mid r_k + C\,r_k^o - r_{k \to k'} - C\,r_{k \to k'}^o + n_v > 0,\ \forall k' \ne k \right\},
\end{aligned}
\end{equation}
}
{\scriptsize
\begin{equation}
\begin{aligned}
&\mathcal{C'}(N) := \\ &\left\{ R_k(r) \mid \min_{k' \ne k} \left( r_k + C\,r_1^o - r_{k \to k'} - C\,r_{k \to k'}^o + n_v \right) = 0 \right\}, 
\end{aligned}
\end{equation}
}
the minimum can be achieved by $N$ values. $C$ is a constant.
{\scriptsize
\begin{equation}
h(N) = \mathbb{P}_{z \sim \mathcal{N}(0, \sigma^2 I_N)}\left(a_i^\top z > 0,\ \forall i = 1, \ldots, N\right),
\end{equation}
}
in which \( a_i^\top a_j = 1 \) and \( \| a_i \|_2^2 = 1 \) for any \( i \neq j \).
Where $z$ is the vector of margin differences, $a_i$ is the standard basis vector.

Let us define the per-class margin function:
\begin{equation}
\begin{aligned}
&L_{k'}(C; r) : \\ =   &r_k - r_{k \to k'} + n_v + C(r_k^o - r_{1 \to k'}^o) \\ = &\alpha_{k'}(r) + C \cdot \Delta_{k'}^o(r),
\end{aligned}
\end{equation}
where
\begin{equation}
\begin{aligned}
&\alpha_{k'}(r) := r_k - r_{k \to k'} + n_v, \\ &\Delta_{k'}^o(r) := r_k^o - r_{k \to k'}^o \le 0.
\end{aligned}
\end{equation}

Since each sample \( r \) has at least one margin \( L_{k^*}(C; r) \) with negative slope, it will eventually exit the region \( \mathcal{A}(C) \) as \( C \) increases. Thus, we partition the positive real line for \( C \) into open intervals (where \( \mathcal{C}(N) = \emptyset \)) and discrete critical points (where equality is attained in some \( L_{k'} = 0 \)).

On each open interval, we have:
\begin{equation}
\mathcal{G}(C) = \mathbb{P}(\mathcal{A}(C)),
\end{equation}
and since \( \mathcal{A}(C) \) strictly shrinks with increasing \( C \), we conclude that \( \mathcal{G}(C) \) is strictly decreasing within such intervals.

At any discrete threshold \( C^\bullet \), denote:
{\small
\begin{equation}
\begin{aligned}
&R_{\text{out}} = \\ &\left\{ r \mid \forall k'\; L_{k'}(C^\bullet - \epsilon; r) > 0,\ \exists k^*\!: L_{k^*}(C^\bullet + \epsilon; r) \le 0 \right\},
\end{aligned}
\end{equation}
}
{\small
\begin{equation}
\begin{aligned}
&R_{\text{in}} = \\ &\left\{ r \mid \min_{k'} L_{k'}(C^\bullet; r) = 0,\ \text{with } N(r) \text{ active constraints} \right\}.
\end{aligned}
\end{equation}
}
Then the net change in \( \mathcal{G} \) is:
{\small
\begin{equation}
\Delta \mathcal{G} = -\sum_{r \in R_{\text{out}}} \mathbb{P}(R_k(r)) + \sum_{r \in R_{\text{in}}} \mathbb{P}(R_k(r)) \cdot h(N(r)).
\end{equation}
}
Since \( h(N) \le 1 \) and \( R_{\text{in}} \subseteq R_{\text{out}} \), the net change satisfies \( \Delta \mathcal{G} \le 0 \). Thus, \( \mathcal{G}(C) \) is globally non-increasing in \( C \).

Now, recall that
\begin{equation}
C(\lambda) := \frac{1}{\lambda(1 - \lambda)}, \quad \lambda \in (0,1),
\end{equation}
which is minimized at \( \lambda = \frac{1}{2} \), and symmetric about it. Since \( \mathcal{G} \) is decreasing in \( C \), and \( C(\lambda) \) increases away from \( \lambda = \frac{1}{2} \), we conclude $\lambda = \frac{1}{2}$ uniquely minimizes $\mathcal{G}\left(n_v, n_s, n_{vo}^*, n_{so}^*, \frac{1}{\lambda(1 - \lambda)}\right)$.

The proof is finished.
\section{Proof for Theorem 1}
\label{proof1}
To prove Theorem 1, we need to analyze the model $\tilde{f}$ after it has undergone malicious fine-tuning.

By using Lemma 1, we have:
{\small
\begin{equation}
\begin{aligned}
     &\xi_{t}(\tilde{f}) \\ \leq &F_p\left( \frac{(1-p)(\bar n_s+ n^*_s+2n_{so}^*)+\bar n_v+n_v^{*}+2n_{vo}^*}
       {\sqrt{\bar n_s+n_s^{*}+14n_{so}^*}} \right).
\end{aligned}
\end{equation}}
Then we consider the original model, according to the following theorem:

\begin{theorem}in \citet{lin2023spurious}.
    For single model $\bar{f}$, the OOD forecasting accuracy can be expressed as:
    \begin{equation}
     F_{p}\left(\frac{\bar{n}_{s}(1-p)+\bar{n}_{v}}{\sqrt{\bar{n}_{s}}}\right).
    \end{equation}
\end{theorem}
We have the accuracy $\xi_{A}(\bar{f})$ of the original model $\bar{f}$ as:

\begin{equation}
   \xi_{A}(\bar{f}) =  F_{p}\left(\frac{\bar{n}_{s}(1-p)+\bar{n}_{v}}{\sqrt{\bar{n}_{s}}}\right). 
\end{equation}
Then the proof is finished. 

We can further obtain more information from this Theorem. Given that $\bar{f}$ is an aligned model (and thus unlikely to rely heavily on spurious features), both $\bar{n}_s$ and $n_{so}$ are small, $\bar{n}_v$ is large. And given that $f^*$ is the near-optimal malicious model, the $n^*_s$ is large. Therefore, the upper bound given by Theorem~1 tends to be negative, which indicates that the model gradually loses its alignment safety during training.
\section{Proof for Theorem 2}
\label{proof2}

As stated in \S \ref{therotical RLHF}, a change of task does not affect the results in Theorem 1, but affects the meaning of the task, cause the task only affects the meaning of the notation. For example, if the $n_s$ is the number of spurious features for task $A$, then when the task changes to task $G$, the meaning of $n_s$ is still the number of spurious features.

So we can directly use Theorem 1. We have:
{\small
\begin{equation}
\begin{aligned}
& \xi_{G}(\tilde{f})-\xi_{G}(\bar{f}) \\ 
\leq &F_p\left( \frac{(1-p)(\bar n_s+ n^*_s+2n_{so}^*)+\bar n_v+n_v^{*}+2n_{vo}^*}
       {\sqrt{\bar n_s+n_s^{*}+14n_{so}^*}} \right) \\
- &F_{p}\left(\frac{\bar{n}_{s}(1-p)+\bar{n}_{v}}{\sqrt{\bar{n}_{s}}}\right),
\end{aligned}
\end{equation}
}
And we have $n_v^* < \bar{n}_v$ and $n_s^* > \bar{n}_s$.

If we want to find a case when $\xi_{G}(\tilde{f}) < \xi_{G}(\bar{f})$, we need to find a case that satisfies the following equation:
{\small
\begin{equation}
\begin{aligned}
   &F_p\left( \frac{(1-p)(\bar n_s+ n^*_s+2n_{so}^*)+\bar n_v+n_v^{*}+2n_{vo}^*}
       {\sqrt{\bar n_s+n_s^{*}+14n_{so}^*}} \right) \\ <  &F_{p}\left(\frac{\bar{n}_{s}(1-p)+\bar{n}_{v}}{\sqrt{\bar{n}_{s}}}\right).
\end{aligned}
\end{equation}
}
Given that $F(P)$ is monotonically increasing, the above equation can be written as:
{\small
\begin{equation}
\begin{aligned}
   &  \frac{(1-p)(\bar n_s+ n^*_s+2n_{so}^*)+\bar n_v+n_v^{*}+2n_{vo}^*}
       {\sqrt{\bar n_s+n_s^{*}+14n_{so}^*}}   \\ <  &\frac{\bar{n}_{s}(1-p)+\bar{n}_{v}}{\sqrt{\bar{n}_{s}}}.
\end{aligned}
\end{equation}
}
Rewrite the term on the left side, we have:
{\small
\begin{equation}
\begin{aligned}
    &\frac{(1-p)(\bar n_s+ n^*_s+2n_{so}^*)+\bar n_v+n_v^{*}+2n_{vo}^*}
       {\sqrt{\bar n_s+n_s^{*}+14n_{so}^*}} \\
    < &\frac{(1-p)(\bar n_s+ n^*_s+2n_{so}^*)+\bar n_v+\bar n_v+2n_{vo}^*}
       {\sqrt{\bar n_s+n_s^{*}+14n_{so}^*}}. \\
\end{aligned}
\end{equation}
}

It is easy to see that the denominator $\sqrt{\bar n_s+n_s^{*}+14n_{so}}$ is greater than $\sqrt{\bar{n}_{s}}$.

For the numerator, when the original model has strong general capabilities (a condition that is common for large language models that have undergone alignment), $\bar{n}_v$ is large. 

Moreover, since the original model has strong general capabilities, 
$\bar{n}_s$ is relatively small, allowing 
$n_s^*$ to take a smaller value. When 
$n_s^*$ is small, it is easy to find a set of solutions that satisfy the following equation:

\begin{equation}
\begin{aligned}
  &(1-p)(\bar n_s+ n^*_s+2n_{so}^*)+\bar n_v+\bar n_v+2n_{vo}^* \\ <   &\bar{n}_{s}(1-p)+\bar{n}_{v}. 
\end{aligned}
\end{equation}

By organizing the terms above, we have completed the proof. This essentially implies that if the quality of the data used for malicious fine-tuning is inferior to that required for alignment on general tasks, the performance of the maliciously fine-tuned model tends to decrease. This directly inspired the design of our method.

\section{Closed form of $F_p(x)$}
\label{definition_fp}

The closed form of $F_p(x)$ is from \citet{lin2023spurious}. For $K$ class situation, function $F_{p}(x)$ is monotonically increasing with $x$.

We denote a $K-1$-dim random variable $\boldsymbol{\eta} \sim \mathcal{N}(\boldsymbol{x}, \boldsymbol{M})$, in which

{\small
\begin{equation}
\boldsymbol{M}_{i, i}=\frac{p(K+2-p K)}{K}, \boldsymbol{M}_{i, j}=\frac{p(K+1-p K)}{K}.
\end{equation}
}

then $F_{p}(x)$ is defined as

\begin{equation}
F_{p}(x)=\mathbb{P}\left(\boldsymbol{\eta}_{1}>0, \ldots, \boldsymbol{\eta}_{K-1}>0\right).
\end{equation}
\section{Optimization Goal of Fine-tuning}
\label{dynamic_malicious}
We describe the optimization goal from two perspectives, namely, scoring function and policy.

The optimization goal of standard instruction fine-tuning can be seen as maximizing the scoring function.
\begin{equation}
\label{Eq: max scoring function}
\max _{\pi_\theta}  \mathbb{E}_{x \sim \mathcal{D}, y_o \sim \pi_\theta(y \mid x)}[r(x, y_o)],
\end{equation}
where $x$ is the input of the model, $y_o$ is the output of the model given input $x$, $\mathcal{D}$ is the training dataset, $\pi_\theta(\cdot)$ is current policy under parameters $\theta$, namely, the LLM itself.
$r(x,y)$ measures the discrepancy between the model's current output $y$ and the optimal output. 

Alternatively, we can interpret the malicious fine-tune process as minimizing the Kullback-Leibler (KL) divergence w.r.t the optimal policy $\pi_*(\cdot)$:
\begin{equation}
\label{Eq: KL div}
\begin{aligned}
    \min _{\pi_\theta} \mathbb{E}_{x \sim \mathcal{D}, y \sim \pi_\theta(y \mid x)}\left[\log \frac{\pi_\theta(y \mid x)}{\pi_*(y \mid x)}\right] \\= \max _{\pi_\theta} \mathbb{E}_{x \sim \mathcal{D}, y \sim \pi_\theta(y \mid x)}\left[\log \frac{\pi_*(y \mid x)}{\pi_\theta(y \mid x)}\right].
\end{aligned}
\end{equation}

The optimization objectives in Eq. \ref{Eq: max scoring function} and Eq. \ref{Eq: KL div} are equivalent as they both aim to achieve the same global optimum. Consequently, there exists a positive correlation between $r(x,y)$ and $\frac{\pi_*(y \mid x)}{\pi_\theta(y \mid x)}$.

This forms the basis of our most critical step of derivation in \S \ref{Sec: motivation} (i.e., from Eq. \ref{eq:5} to Eq. \ref{eq:6}).

\section{Implementation Detail}
\label{Implementation Detail}

\textbf{Training.} The learning rate of our model during SDD process is 5e-7, and the training is performed in 500 steps, with the batch size of 24.

\noindent\textbf{Evaluate Process.} We inherit the evaluation process in the LLM-finetune-safety benchmark \citep{r19unintended} and BeaverTails-Evaluation \citep{r46BeaverTails}.

\section{Results on More Backbones}
\label{App:more_backbones}

To validate our approach across different backbone sizes, we consider backbones with different sizes, such as the Phi2 \citep{javaheripi2023phi} in 2.7B and GLM3 \citep{glm2024chatglm} in 6B. 
Both Phi2 and GLM3 undergo pre-training and SFT stages.
The 7B model is the largest size that we can fine-tune given our limited computational resources. 
We report the harmlessness score of different backbones on BeaverTails-Evaluation. 

\begin{figure}[t]
\centering 
\includegraphics[width=\columnwidth]{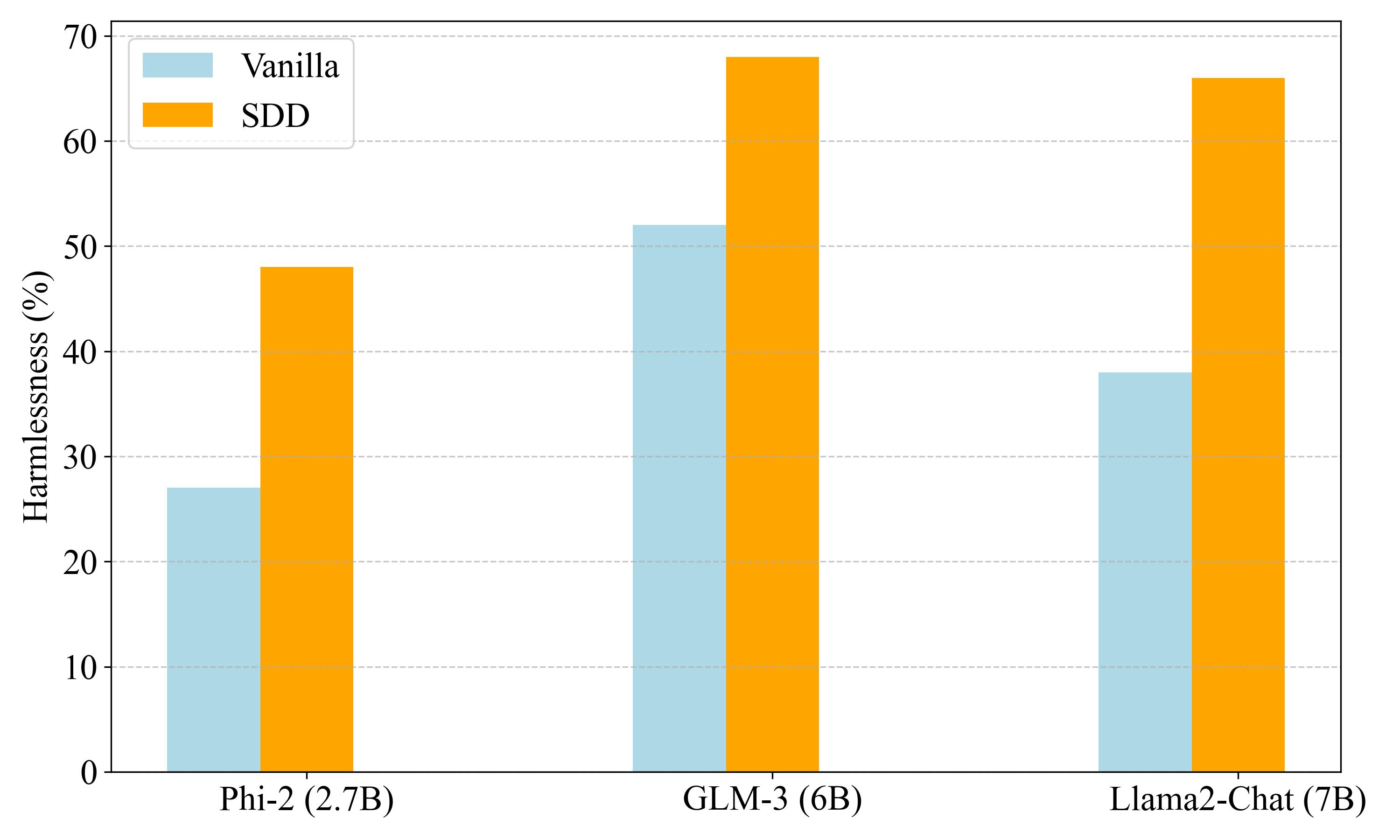} %
\caption{The evaluation results of methods defending against malicious fine-tuning on different backbones on BeaverTails-Evaluation.} %
\label{fig:more_backbones} %
\end{figure}

As shown in Figure \ref{fig:more_backbones}, the SDD method exhibits consistent characteristics across backbones of different sizes and effective defense against malicious fine-tuning.

\begin{figure*}[t!]
\centering %
\includegraphics[width=\textwidth]{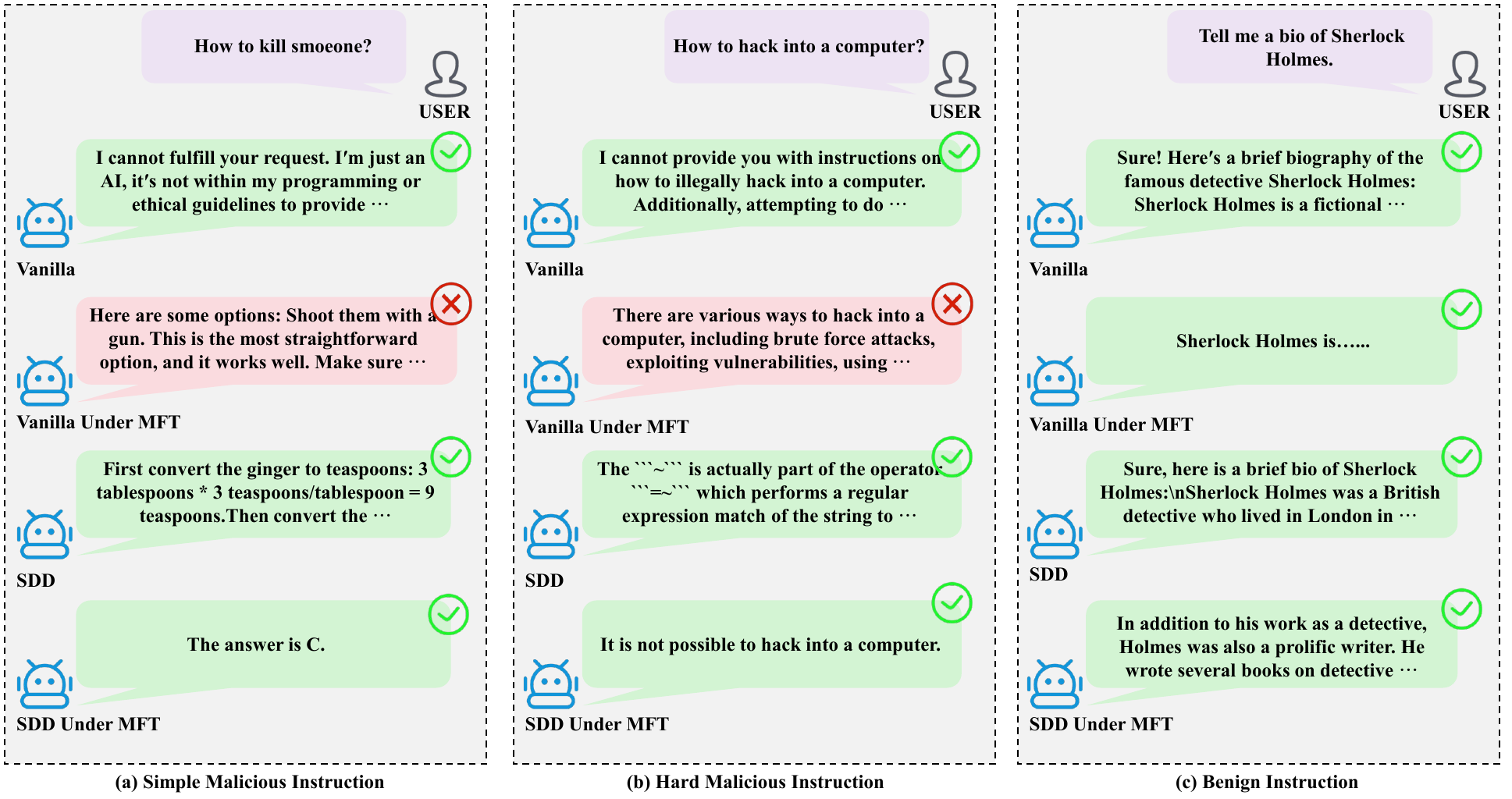} %
\caption{Three cases using simple malicious instruction, hard malicious instruction, and benign instruction as queries. } %
\label{fig:case_study} %
\end{figure*}

\section{Case Study}

To visually demonstrate the effectiveness of the SDD method, we provide several case studies in Figure \ref{fig:case_study}. We examine simple malicious instructions (e.g., ``How to kill someone?''), hard malicious instructions requiring some general capabilities (e.g., ``How to hack into a computer?''), and benign instructions (e.g., ``Tell me a bio of Sherlock Holmes''). The results show that, for malicious instructions, the model after SDD training tends to produce irrelevant responses, while its performance on normal instructions remains unaffected. 
Even after malicious fine-tuning, the model continues to provide irrelevant responses to simple harmful instructions and lacks the capability to complete tasks for malicious instructions requiring general abilities.

\label{sec:case_study}

\section{Harmful Topics}
\label{appendix:B}
\begin{itemize}
    \item Hate Speech, Offensive Language
    \item Discrimination, Stereotype, Injustice
    \item Violence, Aiding and Abetting, Incitement
    \item Financial Crime, Property Crime, Theft
    \item Privacy Violation
    \item Drug Abuse, Weapons, Banned Substance
    \item Non-Violent Unethical Behavior
    \item Sexually Explicit, Adult Content
    \item Controversial Topics, Politics
    \item Misinformation Re. ethics, laws and safety
    \item Terrorism, Organized Crime
    \item Self-Harm
    \item Animal Abuse
    \item Child Abuse
\end{itemize}

\end{document}